\documentclass[english,american,nofootinbib]{elsarticle}
\usepackage[T1]{fontenc}
\usepackage[latin9]{inputenc}
\usepackage{amssymb}

\makeatletter

\providecommand{\tabularnewline}{\\}

\makeatother

\usepackage{babel}
\begin{document}

\title{Generalized hidden Kerr/CFT}

\author{David A. Lowe}

\author{Antun Skanata}

\address{Brown University, Department of Physics, Box 1843, Providence, RI
02912, USA }
\begin{abstract}
We construct a family of vector fields that generate local symmetries
in the solution space of low frequency massless field perturbations
in the general Kerr geometry. This yields a one-parameter family of
$SL(2,\mathbb{R})\times SL(2,\mathbb{R})$ algebras. We identify limits
in which the $SL(2,\mathbb{R})\times SL(2,\mathbb{R})$ algebra contracts
to an $SL(2,\mathbb{R})$ symmetry of the Schwarzschild background.
We note that for a particular value of the free parameter, the symmetry
algebra generates the quasinormal mode spectrum of a Kerr black hole
in the large damping limit, suggesting a connection between the hidden
conformal symmetry and a fundamental CFT underlying the quantum Kerr
black hole. 
\end{abstract}
\maketitle

\section{Introduction}

The hidden Kerr/CFT proposal \citep{Castro2010} has drawn a good
amount of attention since its formulation, and in the past year has
been applied to a number of gravitational backgrounds \citep{Fareghbal2010,Krishnan2010,Cvetic2011,Cvetic:2011dn,Anninos2011}
The allure of the proposal lies in the notion that conformal symmetries
need not be manifest symmetries of the geometry in order to consider
a conformal field theory (CFT) description of low frequency scattering
processes in Kerr background. This should be contrasted with the usual
geometric approach where AdS/CFT methods may be applied for near-extremal
black holes with throat geometries containing AdS subspaces, or for
more general black holes in asymptotically AdS spacetimes.

Low frequency physics in black hole backgrounds has already proved
its fruitfulness in \citep{Maldacena1997} by observing that the scalar
low energy decay spectrum shows characteristic behavior seen in CFT
correlation functions. Similar results have been obtained in \citep{Cvetic1997}
and in the presently discussed case of low frequency massless scalar
excitations in the near region of the Kerr black hole \citep{Castro2010}.
The hope is that by studying this low frequency limit, we learn about
the underlying conformal field theory conjectured to provide a holographic
description of the full quantum Kerr black hole.

In the present work we generalize the hidden CFT generators of \citep{Castro2010}
to a one-parameter family. For special values of the parameter, a
contraction to a single $SL(2,\mathbb{R})$ factor generates symmetries
of the Schwarzschild black hole. Here we find agreement with Schwarzschild
symmetry generators found in \citep{Bertini2011}. Moreover, if we
assume that the $SL(2,\mathbb{R})$ factors are enhanced to full Virasoro
symmetries underlying the CFT, state counting in the CFT is able to
reproduce the exact Kerr entropy. Finally we speculate on the connection
between this hidden CFT and a more fundamental CFT describing the
black hole. In the large damping limit, hints of CFT structure also
emerge \citep{Keshet2008}, and we are able to reproduce the spectrum
of the Kerr quasinormal modes from a particular choice of our free
parameter, generalizing the result of \citep{Bertini2011} for Schwarzschild.

\section{Scalar field equation}

Our analysis will be very much in the spirit of hidden Kerr/CFT proposal
\citep{Castro2010}. The Kerr metric is given in Boyer-Lindquist coordinates:

\begin{eqnarray}
ds^{2} & = & -\left(1-2Mr/\Sigma\right)dt^{2}-\left(4Mar\,\sin^{2}\theta/\Sigma\right)dtd\phi+\left(\Sigma/\Delta\right)dr^{2}+\label{eq:kerrmetric}\\
 &  & \Sigma d\theta^{2}+\sin^{2}\theta\left(\Delta+2Mr(r^{2}+a^{2})/\Sigma\right)d\phi^{2},\nonumber 
\end{eqnarray}
where $\Sigma=r^{2}+a^{2}\cos^{2}\theta$, $\Delta=r^{2}+a^{2}-2Mr$,
$M$ is the black hole mass and $J=Ma$ angular momentum. The two
horizons are located at $r_{\pm}=M\pm\sqrt{M^{2}-a^{2}}$.

A massless scalar field $\psi(t,r,\theta,\phi)$ propagating in such
a background satisfies a Teukolsky \citep{Teukolsky1972} wave equation,
and with the following ansatz%
\footnote{The results below generalize in a straightforward way to massless
spin 1 and spin 2 wave equations, see for example \citep{Lowe2011}.
Here we restrict out attention to the scalar field case.%
}

\[
\psi\sim e^{im\phi-i\omega t}S(\theta)R(r),
\]
the equation can be separated into an angular and radial part:

\begin{eqnarray}
\left[\frac{1}{\sin\theta}\partial_{\theta}\left(\sin\theta\partial_{\theta}\right)-\frac{m^{2}}{\sin^{2}\theta}+\omega^{2}a^{2}\cos^{2}\theta+K_{l}\right]S(\theta) & = & 0,\label{eq:angular}\\
\begin{array}{l}
\left[\partial_{r}\left(\Delta\partial_{r}\right)+\frac{\left(2Mr_{+}\omega-am\right)^{2}}{\left(r-r_{+}\right)\left(r_{+}-r_{-}\right)}-\frac{\left(2Mr_{-}\omega-am\right)^{2}}{\left(r-r_{-}\right)\left(r_{+}-r_{-}\right)}\right.\end{array}\nonumber \\
\left.+(r^{2}+2M(r+2M))\omega^{2}-K_{l}\right]R(r) & = & 0\,.\label{eq:radial}
\end{eqnarray}

In the low frequency limit $M\omega\ll1$, the equation (\ref{eq:angular})
reduces to a Laplacian on $S^{2}$, with eigenfunctions being spherical
harmonics and eigenvalues $K_{l}\approx l(l+1)$, $l=0,1,\ldots$
. We will not focus any further on the angular equation (\ref{eq:angular})
and its properties beyond zeroth order solution; the interested reader
might want to consult \citep{Fackerell1977,Berti2006} and references
within.

For the radial equation (\ref{eq:radial}) we consider a low frequency,
near region limit

\begin{equation}
r\omega\ll1\,,\qquad M\omega\ll1,\label{eq:nearlimit}
\end{equation}
following \citep{Castro2010}. This allows one to drop the $(r^{2}+2M(r+2M))\omega^{2}$
term in (\ref{eq:radial}), at which point the equation reduces to
hypergeometric form. 

If one is interested in Kerr black holes far from extremality, another
interesting possibility arises. Namely, one can demand that $r-r_{-}$
be sufficiently large that the order $\omega$ and higher terms coming
from the pole near $r\to r_{-}$ in (\ref{eq:radial}) be subleading.
That is, we may introduce the deformation parameter $\kappa$ and
deform (\ref{eq:radial}) to

\begin{equation}
\left[\partial_{r}\left(\Delta\partial_{r}\right)+\frac{\left(2Mr_{+}\omega-am\right)^{2}}{\left(r-r_{+}\right)\left(r_{+}-r_{-}\right)}-\frac{\left(2M\kappa r_{+}\omega-am\right)^{2}}{\left(r-r_{-}\right)\left(r_{+}-r_{-}\right)}\right]R(r)=l(l+1)R(r),\label{eq:trunk2}
\end{equation}
This leaves the low frequency limit unchanged, as long as the two
constraints
\begin{eqnarray}
\frac{\kappa M^{2}am\omega}{(r-r_{-})(r_{+}-r_{-})} & \ll & 1,\nonumber \\
\frac{\kappa^{2}M^{4}\omega^{2}}{(r-r_{-})(r_{+}-r_{-})} & \ll & 1,\label{eq:extraconstraint}
\end{eqnarray}
are satisfied. A version of this deformation for the Schwarzschild
black hole was considered in \citep{Bertini2011}. It should be noted
that these conditions are implied by the near-region condition (\ref{eq:nearlimit})
as long as $r-r_{-}$ does not vanish. Thus these conditions are a
rather weak modification of the near-region limit.

Let us conclude this section with some additional motivation for the
deformation. One might consider simply deforming the positions of
the singularities and coefficients in (\ref{eq:trunk2}) in an arbitrary
way, such that the wave equation reduced to (\ref{eq:radial}) as
$\omega\to0$. However if the coefficient involving the singularity
at $r=\infty$ or at $r=r_{+}$ is deformed, the low energy solutions
to the wave equation are changed in a drastic way, since the coefficients
control the divergence of the solutions at the singular points. Shifting
the positions of these singularities produces a deformation that could
only be explained by an action involving more than two time derivatives,
which we choose not to consider in the present work. However the inner
horizon $r=r_{-}$ is a special case, because we expect the full nonlinear
solution for a perturbation of Kerr to become singular there. Including
back-reaction is expected to yield an asymptotically null spacelike
singularity capping the would-be inner horizon \citep{Brady:1998ht}.
Since the low energy linearized wave equation is not relevant at this
singular surface near $r=r_{-}$ it is natural to explore deformations
of the wave equation near this point. The $\kappa$ deformation in
(\ref{eq:trunk2}) is the unique such deformation of the linearized
equation of motion, yielding an equation of motion second order in
time derivatives.

\section{Constructing $SL(2,\mathbb{R})$\label{sec:Constructing}}

We consider a set of vector fields that satisfy an $SL(2,\mathbb{R})$
algebra. We also demand the quadratic Casimir reproduces the scalar
field wave equation in the Kerr background in the near region low
frequency approximation (\ref{eq:nearlimit}) subject to the additional
constraints (\ref{eq:extraconstraint}). One general form of such
vector fields is

\begin{eqnarray}
L_{\pm} & = & e^{\pm\alpha t\pm\beta\phi}\left(g_{\pm}(r)\partial_{r}+h_{\pm}(r)\partial_{\phi}+k_{\pm}(r)\partial_{t}\right),\label{eq:vectfields-1}\\
L_{0} & = & \gamma\partial_{t}+\delta\partial_{\phi}\,.\nonumber 
\end{eqnarray}
The requirement that $L_{0}$ is an eigenvector of a state $\psi\sim e^{im\phi-i\omega t}R(r)S(\theta)$
sets $\gamma$ and $\delta$ to constants. The constraints we impose
are:

\begin{eqnarray*}
\left[L_{+},L_{-}\right] & = & 2L_{0},\\
\left[L_{\pm},L_{0}\right] & = & \pm L_{\pm},\\
L_{0}^{2}-\frac{1}{2}(L_{+}L_{-}+L_{-}L_{+}) & = & \partial_{r}(\Delta\partial_{r})+f(r)\,,
\end{eqnarray*}
where $f(r)$ is a function that involves no single derivatives in
$t$ or $\phi$, except for $\partial_{t}\partial_{\phi}$. Given
these constraints, we claim the following is the most general functional
form of such generators: 

\begin{eqnarray}
L_{\pm} & = & e^{\pm\alpha t\pm\beta\phi}\left(\mp\sqrt{\Delta}\partial_{r}+\frac{C_{2}-\delta r}{\sqrt{\Delta}}\partial_{\phi}+\frac{C_{1}-\gamma r}{\sqrt{\Delta}}\partial_{t}\right),\label{eq:generators-1}\\
L_{0} & = & \gamma\partial_{t}+\delta\partial_{\phi}\,,\nonumber 
\end{eqnarray}
with constraints on parameters arising from imposing the $sl(2,\mathbb{R})$
algebra:

\begin{eqnarray}
\alpha C_{1}+\beta C_{2}+M & = & 0,\label{eq:constraints}\\
1+\alpha\gamma+\beta\delta & = & 0.\nonumber 
\end{eqnarray}
These determine $\alpha$ and $\beta$. The last three equations we
impose are the ones identifying appropriate terms in the quadratic
Casimir with $\partial_{\phi}^{2}$, $\partial_{t}\partial_{\phi}$
and $\partial_{t}^{2}$ terms in the wave equation (\ref{eq:trunk2}).
The $\partial_{\phi}^{2}$ term gives us the branches:

\[
\begin{array}{lcl}
\delta=\pm a/\sqrt{M^{2}-a^{2}} & \,\,\,\,\, & \delta=0\\
C_{2}=M\delta & \,\,\,\,\, & C_{2}=\pm a\,.
\end{array}
\]
The differing signs simply generate automorphisms of the algebra,
so may be dropped in the following. Examining the two remaining terms
gives:

\[
\gamma\delta a^{2}-C_{1}C_{2}-r(2M\gamma\delta-C_{2}\gamma-C_{1}\delta)=-\frac{2Mr_{+}a}{r_{+}-r_{-}}\left[r(1-\kappa)-(r_{-}-\kappa r_{+})\right],
\]
and

\[
\gamma^{2}a^{2}-C_{1}^{2}-2r\gamma(M\gamma-C_{1})=-\frac{4M^{2}r_{+}^{2}}{r_{+}-r_{-}}\left[r(1-\kappa^{2})-(r_{-}-\kappa^{2}r_{+})\right]\,.
\]
The two possible branches are shown in Table \ref{tab:Two-branches-of}
describing a one-parameter family of $SL(2,\mathbb{R})\times SL(2,\mathbb{R})$
generators labelled by $\kappa$.

\begin{table}
\begin{tabular}{|c|c|}
\hline 
\selectlanguage{english}%
$\gamma=\frac{2Mr_{+}}{r_{+}-r_{-}}(\kappa+1)$\selectlanguage{american}%
 & \selectlanguage{english}%
$\gamma=\frac{2Mr_{+}}{r_{+}-r_{-}}(\kappa-1)$\selectlanguage{american}%
\tabularnewline
\selectlanguage{english}%
$\delta=\frac{2a}{r_{+}-r_{-}}$\selectlanguage{american}%
 & \selectlanguage{english}%
$\delta=0$\selectlanguage{american}%
\tabularnewline
\selectlanguage{english}%
$C_{1}=\frac{2Mr_{+}}{r_{+}-r_{-}}(\kappa r_{+}+r_{-})$\selectlanguage{american}%
 & \selectlanguage{english}%
$C_{1}=\frac{2Mr_{+}}{r_{+}-r_{-}}(\kappa r_{+}-r_{-})$\selectlanguage{american}%
\tabularnewline
\selectlanguage{english}%
$C_{2}=M\delta$\selectlanguage{american}%
 & \selectlanguage{english}%
$C_{2}=a$\selectlanguage{american}%
\tabularnewline
\hline 
\end{tabular}

\caption{\label{tab:Two-branches-of}Two branches of solutions for the generators.}
\end{table}
 The solution for the generators is

\begin{eqnarray}
L_{\pm} & = & \begin{array}{r}
e^{\mp2\pi T_{R}\phi}\left[\mp\sqrt{\Delta}\partial_{r}-\frac{1}{2\pi T_{H}}\frac{r-M}{\sqrt{\Delta}}\left(\Omega\partial_{\phi}+\partial_{t}\right)\right.\\
\left.+\frac{1}{2\pi\Omega(T_{L}+T_{R})}\frac{r-r_{+}}{\sqrt{\Delta}}\partial_{t}\right]
\end{array}\label{eq:L-1}\\
L_{0} & = & \frac{1}{2\pi T_{H}}\left(\Omega\partial_{\phi}+\partial_{t}\right)-\frac{1}{2\pi\Omega(T_{L}+T_{R})}\partial_{t}\nonumber 
\end{eqnarray}
and

\begin{eqnarray}
\bar{L}_{\pm} & = & \begin{array}{r}
e^{\pm2\pi\Omega(T_{L}+T_{R})t\mp2\pi T_{L}\phi}\left[\mp\sqrt{\Delta}\partial_{r}+\frac{2Mr_{+}}{\sqrt{\Delta}}\left(\Omega\partial_{\phi}+\partial_{t}\right)\right.\\
\left.+\frac{1}{2\pi\Omega(T_{L}+T_{R})}\frac{r-r_{+}}{\sqrt{\Delta}}\partial_{t}\right],
\end{array}\label{eq:Lbar-1}\\
\bar{L}_{0} & = & -\frac{1}{2\pi\Omega(T_{L}+T_{R})}\partial_{t},\nonumber 
\end{eqnarray}
where $T_{H}=\frac{\sqrt{M^{2}-a^{2}}}{4\pi Mr_{+}}$ is Hawking temperature
and $\Omega=\frac{a}{2Mr_{+}}$ angular velocity at the outer horizon,
and we have introduced ``CFT'' temperatures as

\begin{equation}
T_{R}=\frac{\sqrt{M^{2}-a^{2}}}{2\pi a}\label{eq:righttemp}
\end{equation}

and

\begin{equation}
T_{L}=T_{R}\frac{1+\kappa}{1-\kappa}.\label{eq:lefttemp}
\end{equation}

We note the generators (\ref{eq:Lbar-1}) shift the frequency of a
mode by the imaginary amount $2\pi\Omega(T_{L}+T_{R})$. This means
we only stay within the low frequency limit (\ref{eq:nearlimit})
provided we are close to the extremal Kerr limit, so that $T_{R}\ll1$
(assuming $\kappa$ is fixed). Outside this limit the descendants
of some primary operator are no longer mapped to eigenfunctions of
the low frequency scalar field equation. The other set of generators
(\ref{eq:L-1}) do not suffer from this additional constraint. 

Higher order corrections to the Teukolsky equation in the low frequency
limit give rise to soft breaking of the conformal symmetry, and running
of the anomalous dimensions, as described in \citep{Lowe2011}.

In addition, global identifications on the solution space by $\phi\to\phi+2\pi$
explicitly breaks the symmetry algebra down to $U(1)\times U(1)$
generated by $(L_{0},\bar{L}_{0}).$

Finally let us comment on two special cases where the general solutions
(\ref{eq:L-1}) and (\ref{eq:Lbar-1}) do not hold. For $\kappa=1$
and general rotation parameter $a$, the right branch in Table \ref{tab:Two-branches-of}
fails to yield a consistent solution to the constraint equations,
so only the left branch generates an $SL(2,\mathbb{R})$. Since we
only find one $SL(2,\mathbb{R})$ for general $a$ we are unable to
carry through the conjecture that the theory should be dual to a 2d
CFT, so we do not pursue this case further in this paper.

The Schwarzschild case, $a=0$, should likewise be treated as a special
case. Here we find the constraint equations are inconsistent unless
$\kappa=\pm1$. For both values of $\kappa$ the same single copy
of $ $$SL(2,\mathbb{R})$ is found. This may be read off, for example,
from the right branch of Table \ref{tab:Two-branches-of} by setting
$\kappa=-1$ and $a=0$.

\section{Relation to known results}

If we set $\kappa=r_{-}/r_{+}$ our results match those of \citep{Castro2010}.
For this choice of $\kappa$ the pole term in (\ref{eq:radial}) is
exact, so the subsidiary constraints (\ref{eq:extraconstraint}) may
be dropped. 

It is also interesting to consider the Schwarzschild limit, where
our results match those of Bertini, Cacciatori and Klemm \citep{Bertini2011}.
They find a single set of $SL(2,\mathbb{R})_{Sch}$ generators
\begin{equation}
L_{\pm}=e^{\pm t/4M}\left(\pm\sqrt{\Delta}\partial_{r}-\frac{4M(r-M)}{\sqrt{\Delta}}\partial_{t}\right),L_{0}=-4M\partial_{t}.\label{eq:klemm}
\end{equation}
The generators constructed at the end of Section \ref{sec:Constructing}
for $a=0$ and $\kappa=-1$ match these, up to the automorphism $L_{\pm}\to-L_{\pm}$
and $L_{0}\to L_{0}.$

\section{Quasinormal modes}

It is well known that classical black holes are characterized by a
discrete set of complex frequencies, named quasinormal modes. The
quasinormal modes correspond to a certain set of boundary conditions,
with waves purely outgoing at infinity and ingoing at the horizon.
Two observations immediately jump to mind: a quantum theory of gravity
should reproduce this spectrum; and if the states in this quantum
theory are fully characterized by quasinormal modes, studying semiclassical
physics outside the black hole horizon should teach us about this
quantum theory. 

The connection between quasinormal modes for three-dimensional black
holes and CFT states has been made precise in \citep{Birmingham2002}
by looking at linearized perturbations%
\footnote{The previous definition of quasinormal modes via ingoing flux at infinity
does not make it if we put the system in a box. The way quasinormal
modes were defined in asymptotically AdS backgrounds was to impose
either Dirichlet boundary conditions at asymptotic infinity, or a
vanishing flux $\mathcal{F}\sim\sqrt{-g}\left(R^{*}\partial_{\mu}R-c.c.\right)$.
Both choices lead to same spectrum.%
} and showing an explicit agreement between quasinormal frequencies
and the poles of the retarded correlation function in the CFT. 

Likewise for the Kerr black hole, there is a discrete spectrum of
quasinormal modes \citep{Berti2004}. At large damping the imaginary
part of the frequency increases approximately linearly with mode number,
while the real part approaches a constant. According to Keshet and
Neitzke \citep{Keshet2008}, large damping is where one expects a
CFT description to emerge, as the transmission and reflection amplitudes
take a familiar CFT-like form. Moreover, a step towards this understanding
has been made in \citep{Keshet2007}, by obtaining the quasinormal
mode spectrum via a WKB approximation to the wave equation.

An interesting observation was made in \citep{Bertini2011} -- the
descendant states $\left(L_{-}\right)^{n}\psi(t,r,\phi)$ reproduce
the large damping quasinormal spectrum of the Schwarzschild black
hole. We speculate this might be the case with generators (\ref{eq:Lbar-1})
as well, giving a connection between the low frequency hidden CFT,
and some more fundamental underlying CFT that correctly describes
the quasinormal modes.

Keshet and Hod \citep{Keshet2007} compute the quasinormal mode spectrum
at large damping and obtain 

\begin{equation}
\omega=-m\hat{\omega}-2\pi iT_{0}(n+1/2),\label{eq:spectrum}
\end{equation}
to leading order in $n$, where $T_{0}=f(a/M)T_{H}$ is a smooth slowly
varying function of angular momentum with $f(0)=1$, that may be expressed
in general using elliptic integrals.

We can choose the value of $\kappa(a)$ by solving
\[
T_{0}=\Omega T_{R}\frac{2}{1-\kappa}\,.
\]
Then by defining the lowest weight state via

\begin{eqnarray*}
\bar{L}_{0}\Phi^{(0)} & = & \bar{h}\Phi^{(0)},\\
\bar{L}_{+}\Phi^{(0)} & = & 0,
\end{eqnarray*}
it is easy to check the descendants $\Phi^{(n)}=\left(\bar{L}_{-}\right)^{n}\Phi^{(0)}$
reproduce the large $n$ behavior of the spectrum (\ref{eq:spectrum}).

\section{Entropy}

Following the discussion of the introduction, we can propose that
the $SL(2,\mathbb{R})\times SL(2,\mathbb{R})$ symmetry is promoted
to a full left and right-moving Virasoro symmetry in the full quantum
theory. The Cardy formula gives
\begin{equation}
S=\frac{\pi^{2}}{3}\left(c_{L}T_{L}+c_{R}T_{R}\right)\,.\label{eq:cardy}
\end{equation}
The $T_{L}$ and $T_{R}$ appearing in (\ref{eq:righttemp}) and (\ref{eq:lefttemp})
may be matched with the left and right-moving CFT temperatures. The
CFT inherits periodic identifications in the imaginary left and right
moving directions, from the periodic identification of $\phi$, and
the action of the Virasoro generators.

The Bekenstein-Hawking entropy is 
\begin{equation}
S_{BH}=2\pi Mr_{+},\label{eq:bekenstein}
\end{equation}
which agrees exactly with (\ref{eq:cardy}) provided we identify the
central charge as
\begin{equation}
c_{L}=c_{R}=\frac{6a(1-\kappa)Mr_{+}}{\sqrt{M^{2}-a^{2}}}\,.\label{eq:centralcharge}
\end{equation}
We emphasize the identification of the central charge (\ref{eq:centralcharge})
is not an independent computation. Such a computation may well be
possible, but would require a reworking of the Brown-Henneaux calculation
\citep{Brown:1986nw} within the hidden CFT/low frequency framework. 

The central charge depends in a nontrivial way on the deformation
parameter $\kappa$. This indicates the dual description is actually
a family of conformal field theories with a conformal deformation
parameter. While we do not yet have enough data to specify the CFTs
at hand in detail, there are many examples of such families of conformal
field theories. A simple example is the level number of a conformal
field theory based on an affine Lie algebra. A much more general set
of examples, including continuous deformations of CFTs that change
the central charge, appears in \citep{Freericks:1988zg}.

The formula (\ref{eq:centralcharge}) reduces to the result of \citep{Castro2010}
when $\kappa=r_{-}/r_{+}$ where $c_{L}=c_{R}=12J$. There one can
argue that in the extremal limit, the central charge follows from
standard geometric argument \citep{Guica2009a}. However a closer
look at this shows the low frequency symmetry generators of \citep{Castro2010}
do not smoothly match the isometry generators of \citep{Guica2009a}.
Thus the fixed point of the low frequency conformal symmetry of \citep{Castro2010}
does not coincide with extremal Kerr \citep{Lowe2011}. Moreover a
strong argument for the non-renormalization of the central charges
away from the extremal point is lacking for the case considered in
\citep{Castro2010}. The addition of the extra parameter $\kappa$
in (\ref{eq:centralcharge}) does not help this situation.

In the Schwarzschild limit we only retain a single $SL(2,\mathbb{R})$
symmetry, which might be associated with a conformal quantum mechanics.
If we assume this alone is promoted to a full Virasoro symmetry, we
find periodicity with respect to $\phi$ no longer fixes the CFT temperature.
Rather we must return to the generators (\ref{eq:klemm}), where we
can read off the temperature 
\[
T_{CFT}=T_{H}=\frac{1}{8\pi M}\,.
\]
The central charge for conformal quantum mechanics dual to Schwarzschild
is then predicted to be
\begin{equation}
c_{Sch}=96M^{3}\,,\label{eq:schcentralcharge}
\end{equation}
an apparently new result.

\section{Conclusions}

We have obtained a generalization of the hidden low frequency limit
of the wave equation in a general Kerr background. This hints at an
underlying conformal field theory description, that we have explored.
The most exciting aspect of this work is the idea that the underlying
hidden conformal field theory description may be studied via low frequency
scattering, a kind of nonlocal probe of the geometry, rather than
simply looking for geometric isometries as has been already much studied
in the literature. This opens up the possibility that even the Schwarzschild
black hole may have a CFT dual.

To put these results on a firmer footing, it would be interesting
to obtain the central extensions (\ref{eq:centralcharge}) and (\ref{eq:schcentralcharge})
from some generalization of the asymptotic symmetry algebra analysis
of \citep{Brown:1986nw}. We hope to explore this issue in future
work.

This research is supported in part by DOE grant DE-FG02-91ER40688-Task
A and an FQXi grant.

\bibliographystyle{elsarticle-num}
\bibliography{hiddenKerrbiblio}

\end{document}